\documentclass{emulateapj}

 \usepackage{graphics}
\usepackage{epsfig}
\usepackage{natbib}
\shorttitle{H$_{2}$ Gas in AU Mic}
\shortauthors{France et al.}

\begin{document}


\title{A Low-Mass H$_{2}$ Component to the AU Microscopii Circumstellar Disk\altaffilmark{1}}


\author{Kevin France\altaffilmark{2}, Aki Roberge\altaffilmark{3}, Roxana E. Lupu\altaffilmark{4},
	Seth Redfield\altaffilmark{5}$^{,}$\altaffilmark{6} and Paul D. Feldman\altaffilmark{4}}



    

\altaffiltext{1}{Based in part on observations made with the the $Far$ $Ultraviolet$
 $Spectroscopic$ 
$Explorer$, operated by the Johns Hopkins University for NASA.}

\altaffiltext{2}{Canadian Institute for Theoretical Astrophysics, University of Toronto,
    60 St. George Street, Toronto, ON M5S 3H8; france@cita.utoronto.ca}
\altaffiltext{3}{Exoplanets and Stellar Astrophysics Laboratory, 
	NASA Goddard Space Flight Center, Greenbelt, MD 20771; akir@milkyway.gsfc.nasa.gov}
\altaffiltext{4}{Department of Physics and Astronomy, Johns Hopkins University,
   Baltimore, MD 21218; roxana@pha.jhu.edu, pdf@pha.jhu.edu}
\altaffiltext{5}{Department of Astronomy, University of Texas, Austin, TX 78712; sredfield@astro.as.utexas.edu}
\altaffiltext{6}{Hubble Fellow}

\received{{\it March 25, 2007}}
\revised{ {\it May 08, 2007} }
\accepted{{\it July 03, 2007} }



\begin{abstract}
We present a determination of the molecular gas mass in the AU Microscopii
circumstellar disk.  Direct detection of a gas component to the AU Mic disk has
proven elusive, with upper limits derived from ultraviolet absorption line
and submillimeter CO emission studies.  Fluorescent emission lines
of H$_{2}$, pumped by the \ion{O}{6} $\lambda$1032 resonance line
through the $C$--$X$ (1~--~1) $Q(3)$ $\lambda$1031.87 \AA\ transition, 
are detected by the $Far$~$Ultraviolet$~$Spectroscopic$~$Explorer$.
These lines are used to derive the H$_{2}$ column density associated with the AU Mic system.  
The derived column density is in the range 
$N($H$_{2}$$)$~=~1.9 $\times$ 10$^{17}$~--~2.8 $\times$ 10$^{15}$ cm$^{-2}$, roughly 
two orders of magnitude lower than the upper limit inferred from absorption line studies.
This range of column densities reflects the range of H$_{2}$ excitation temperature 
consistent with the observations, $T($H$_{2}$$)$~=~800~--~2000 K, derived from the presence
of emission lines excited by \ion{O}{6} in the absence of those excited by Ly$\alpha$.
Within the observational uncertainties, the data are consistent with the H$_{2}$ gas 
residing in the disk.
The inferred $N($H$_{2}$$)$ range corresponds to H$_{2}$-to-dust ratios of $\lesssim$~$\frac{1}{30}$:1
and a total $M($H$_{2}$$)$ = 4.0~$\times$~10$^{-4}$~--~5.8~$\times$~10$^{-6}$ $M_{\oplus}$. 
We use these results to predict the intensity of the 
associated rovibrational emission lines of H$_{2}$ at infrared wavelengths covered
by ground-based instruments, $HST$-NICMOS, and the $Spitzer$-IRS. 

\end{abstract}



\section{Introduction}

Circumstellar (CS) disks around young main sequence stars appear to be in 
transition between massive, gas-rich protoplanetary disks and low-mass, 
gas-poor planetary systems. 
Surveys for CS disks in young stellar clusters suggest that gas-rich protoplanetary disks dissipate on timescales of 
$\sim$ 1~--~10 Myr~\citep{bally98, haisch01b}. 
For solar-type stars, the timescale for gas dissipation is roughly equal to 
the theoretical time required for gas giant planet formation by the standard core-accretion method \citep[e.g.][]{hubickyj05}.
For low-mass M-stars, on the other hand, the time required to form giant planets by core-accretion is much longer than it is around solar-type stars, and may be longer than the typical gas disk lifetimes \citep{laughlin04}. 
Consequently, observations of the gas component in the disks of low- and intermediate-mass main sequence stars undergoing the transition from gas-rich protoplanetary disk to gas-poor debris disk are important for constraining giant planet formation scenarios.
 
AU Microscopii is a nearby ($d \approx 10$~pc) M1 star surrounded by an 
edge-on (inclination = 1~--~3$^{\circ}$) dust disk \citep{kalas04, krist05}.
The star is a very active flare star, and the majority of AU~Mic 
studies prior to imaging of its disk focused on the stellar activity 
\citep[][and references therein]{redfield02}.
AU~Mic is a member of the $\beta$~Pictoris moving group, indicating an age of 
$t_{AU Mic}$ = 12$^{+8}_{-4}$ Myr~\citep{zuckerman01}.  
$\beta$~Pic (A5V) itself has a well-studied edge-on debris disk 
\citep{roberge00,etangs01}. 
These two stars allow us to examine possible differences between disks of the same age around stars of very different mass. 

\citet{liu04} measured the 850 $\mu$m dust emission from the AU~Mic disk, and inferred a total mass of 0.011~$M_{\oplus}$ at $T_{dust}$ = 40 K.  
No sub-mm CO emission was detected, implying a low CS gas mass; however, this upper 
limit was not very stringent. 
A much lower upper limit on the column density of molecular gas was determined 
using far-UV H$_{2}$ absorption spectroscopy~\citep{roberge05}.
This study indicated that the H$_2$-to-dust ratio in the disk is less than 
about 6:1, dramatically depleted from the canonical interstellar ratio of 100:1. 
\citet{roberge05} mentioned that there was an indication of weak H$_2$ absorption 
at a column density at least an order of magnitude below their upper limit.
They also noted the presence of very weak fluorescent H$_{2}$ emission 
lines in the far-UV spectra, which had been seen in previous work~\citep{redfield02}.

An M-star has insufficient continuum flux to excite (``pump'') far-UV fluorescent emission.  
In general, only stars of type $\sim$B3 and earlier have the 
necessary spectral energy distribution ($\lambda$~$\leq$ 1110 \AA) to produce detectable levels of continuum pumped UV fluorescence 
\citep{kf_thesis}. 
However, pre-main-sequence and low-mass dwarf stars show high-temperature emission lines arising in their chromospheric and coronal regions
\citep{linsky94, wood97, redfield02,redfield03}.
These stellar emission lines can coincide in wavelength with transitions of 
H$_{2}$, providing the necessary flux to excite detectable fluorescence when 
CS material is present.  
Line pumped H$_2$ fluorescence has been observed from a variety of 
environments, including T~Tauri disks \citep{valenti00, wilkinson02, herczeg06},
sunspots \citep{jordan77}, Solar System comet comae \citep{feldman02}, and recently from planetary nebulae \citep{lupu06}. 

Here we present a detailed description and analysis of the fluorescent 
H$_2$ emission lines in \textit{Far Ultraviolet Spectroscopic Explorer} 
(\textit{FUSE}) spectra of AU~Mic.  
The majority of line pumped fluorescence studies have focused
on H$_{2}$ excitation by Ly$\alpha$ emission \citep{shull78, black87}. 
However, in the case of AU~Mic, we show that the observed H$_2$ emission lines are pumped 
by the \ion{O}{6} $\lambda1032$ stellar emission line, which is an important excitation source in some situations \citep{wilkinson02, redfield02, herczeg06}. 
In \S2 of this paper, we briefly describe the observations and the characteristics of the detected fluorescent H$_2$ emission lines. 
In \S3, we calculate the total fluorescent H$_2$ flux, 
constrain the H$_2$ temperature, and 
determine the total column density of absorbing H$_2$. 
Our estimation of the total mass of H$_2$ gas in the AU~Mic system appears 
in \S4.  This section also contains a discussion of the H$_2$ heating and its 
implications for the gas location.
A comparison between the observed AU Mic and $\beta$ Pic disk properties is 
given in \S5, and predictions for the near- and mid-IR H$_2$ emission line fluxes 
of AU Mic appear in \S6.  Our results are summarized in \S7.

\section{$FUSE$ Observations and H$_{2}$ Line Identification}

$FUSE$ performs medium-resolution ($\Delta$v $\approx$ 15 km s$^{-1}$)
spectroscopy in the far-UV bandpass (905~--~1187 \AA).  The $FUSE$ observatory is 
described in Moos et al. (2000) and on-orbit performance characteristics are given in 
Sahnow et al. (2000).
AU Mic was observed with $FUSE$ early in the mission as part of the ``cool stars''
programs P118 (2000 August 26, exposure time = 17.3 ks) and P218 
(2001 October 10, exposure time = 26.5 ks).  These observations are described
in ~\citet{redfield02}.  These data were acquired in TTAG mode through the 
(30\arcsec\ $\times$ 30\arcsec) LWRS aperture.  The data have 
been reprocessed with the CalFUSE pipeline v2.4.0. 
Data taken during periods of stellar flare activity were excluded~\citep{roberge05}.
The fluorescent emission lines studied here fall on the LiF 2A (1087~--~1179 \AA)
and LiF 1B (1092~--~1187 \AA) detector segments.  Archival $HST$ Space Telescope Imaging
Spectrograph (STIS) spectra of the chromospheric \ion{C}{3} $\lambda$1176 multiplet
were used to establish the wavelength calibration of the $FUSE$ LiF 1B channel~\citep{roberge05}.
Due to a lower background in the (1~--~3) $Q(3)$ $\lambda$1119.08
\AA\ region, we present the LiF 1B data here.  ~\nocite{moos00,sahnow00}

\begin{figure}
\begin{center}
\hspace{+0.0in}
\includegraphics[width=2.1in,angle=90]{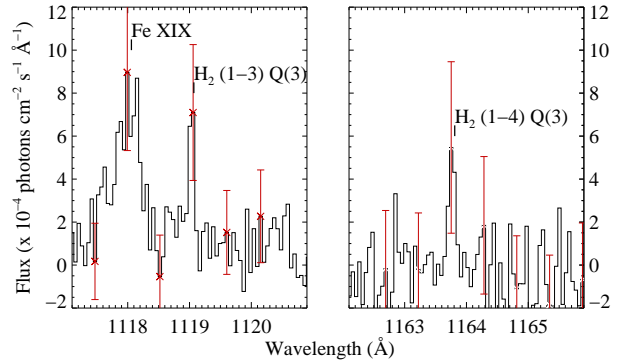}
\caption{\label{fuseobs} $FUSE$ LiF 1B data for the $C$~--~$X$ (1~--~3) $Q(3)$
1119.08 \AA\ and (1~--~4) $Q(3)$ 1163.81 \AA\ emission lines.  The broad feature in the left 
panel is coronal \ion{Fe}{19}~\citep{redfield02,redfield03}.  
For display, measurement uncertainties on the flux calibration are shown
as the red error bars on the x-symbols.}
\end{center}
\end{figure}

The H$_{2}$ emission lines observed in the LiF 1B channel are shown in Figure~\ref{fuseobs}.  The $C$--$X$ (1~--~3) $Q(3)$ $\lambda$1119.077 \AA\ and 
$C$--$X$ (1~--~4) $Q(3)$ $\lambda$1163.807 \AA\ transitions are shown in the 
left and right panels, respectively.  These lines are excited by absorption in the
$C$--$X$ (1~--~1) $Q(3)$ $\lambda$1031.87 \AA\ line which is coincident with the 
strong stellar \ion{O}{6} $\lambda$1032 emission line.  A description of the 
electronic excitation of H$_{2}$ and the molecular notation are given in~\citet{shull82}.
The (1~--~3) and (1~--~4)
lines are expected to be the brightest ones in the \ion{O}{6} pumped fluorescent cascade~\citep{abgrall93b}.  
A Gaussian least-squares fitting routine was used to derive the line strength, width, velocity, and 
detection levels.  The lines were marginally resolved and displayed a slight blueshift with respect
to the stellar velocity (v$_{*}$~=~-4.98 $\pm$ 0.02 km s$^{-1}$).  
The (1~--~3) line was detected at 2.6 $\sigma$ with an integrated line strength
of $I_{(1-3)}$~=~1.61 $\pm$ 0.62 $\times$ 10$^{-15}$ ergs s$^{-1}$ cm$^{-2}$ 
(9.07 $\pm$ 3.49 $\times$ 10$^{-5}$ photons s$^{-1}$ cm$^{-2}$)
in the (30\arcsec~$\times$~30\arcsec) aperture.  Its velocity\footnote{
A detailed description of the $FUSE$ wavelength calibration can be found at:
{\tt http://fuse.pha.jhu.edu/analysis/calfuse.html}.}
was v$_{(1-3)}$~=~-12.0~$\pm$~2.8 km s$^{-1}$ with a FWHM$_{(1-3)}$~=~26~$\pm$~7 km s$^{-1}$
and a peak observed flux of 12.5 $\times$ 10$^{-15}$ ergs s$^{-1}$ cm$^{-2}$ \AA$^{-1}$.
The (1~--~4) line was detected at  1.9 $\sigma$ with an integrated line strength of 
$I_{(1-4)}$~=~1.48 $\pm$ 0.76 $\times$ 10$^{-15}$ ergs s$^{-1}$ cm$^{-2}$ 
(8.67 $\pm$ 4.45 $\times$ 10$^{-5}$ photons s$^{-1}$ cm$^{-2}$) over the same area.
The (1~--~4) emission line was at a velocity
of v$_{(1-4)}$~=~-22.3~$\pm$~4.1 km s$^{-1}$ with a FWHM$_{(1-4)}$~=~35~$\pm$~9 km s$^{-1}$
and a peak observed flux of 9.5 $\times$ 10$^{-15}$ ergs s$^{-1}$ cm$^{-2}$ \AA$^{-1}$.
These findings are summarized in Table 1. 

The observed ratios of the detected emission lines  ($R_{34}$)
are consistent with the theoretical branching ratios within the 1 $\sigma$ uncertainty
($R_{34}$ observed = 1.05$\pm$0.41 vs $R_{34}$ predicted = 0.76). 
The difference between the observed and theoretical ratio may be due to suppression of the 
(1~--~4) 1163.81~\AA\ line by ``the worm'' seen in $FUSE$ spectra from the LiF 1B
channel\footnote{ More information about ``the 
worm'' can be found on the $FUSE$ data analysis page: {\tt http://fuse.pha.jhu.edu/analysis/calfuse\_wp6.html} }.
The line ratios derived from the LiF 2A channel are somewhat closer to the theoretical value, 
although with larger uncertainties ($R_{34}$ observed in LiF 2A = 0.89$\pm$0.55).
The discrepancy is not significant within the observational uncertainties and does not
affect the results presented in Sections 3 and 4.  
The only other H$_{2}$ emission line with a 
comparable branching ratio in the $FUSE$ bandpass is the $C$--$X$ (1~--~0) $Q(3)$ $\lambda$989.73 \AA\
transition.  This wavelength region has a lower instrumental effective area 
and is dominated by a combination of geocoronal \ion{O}{1} and 
stellar \ion{N}{3} lines. No H$_{2}$ emission from $C$--$X$ (1~--~0) $Q(3)$ was observed.

\begin{deluxetable}{ccc}
\tabletypesize{\small}
\tablecaption{H$_{2}$ Emission Line Parameters from the $FUSE$ Observations.
\label{UVcols}}
\tablewidth{0pt}
\tablehead{
\colhead{Line Parameter} & \colhead{H$_{2}$~(1~--~3) $Q(3)$} 
& \colhead{H$_{2}$~(1~--~4) $Q(3)$} \\ 
}
\startdata
$\lambda_{o}$ &  1119.077 \AA\ & 1163.807 \AA\ \\
$\lambda_{obs}$  & 1119.03 \AA\   & 1163.72 \AA\  \\
v  & -12.0 $\pm$ 2.8 km s$^{-1}$ & -22.3 $\pm$ 4.1 km s$^{-1}$ \\
FWHM & 26 $\pm$ 7 km s$^{-1}$& 35 $\pm$ 9 km s$^{-1}$ \\
$I_{ik}$\tablenotemark{a} 	& 9.07 $\pm$ 3.49 $\times$ 10$^{-5}$	& 8.67 $\pm$ 4.45 $\times$ 10$^{-5}$ \\
$F_{peak}$\tablenotemark{b} 	& 7.0 $\times$ 10$^{-4}$	& 5.6 $\times$ 10$^{-4}$\\
 \enddata

\tablenotetext{a}{Integrated Line Strength (photons cm$^{-2}$ s$^{-1}$)}
\tablenotetext{b}{Peak Line Strength (photons cm$^{-2}$ s$^{-1}$ \AA$^{-1}$)}

\end{deluxetable}

The H$_{2}$~$C$--$X$ (1~--~5) $Q(3)$ $\lambda$1208.93~\AA\ and $C$--$X$ (1~--~6) $Q(3)$
$\lambda$1254.11 \AA\ emission lines located in the STIS bandpass have branching ratios  
relative to (1~--~3) 1119.08 \AA\ ($R_{53}$ and $R_{63}$) of roughly 0.6 and 0.1, 
respectively.  Emission lines were detected at these wavelengths at the 2~--~3~$\sigma$
level.  Their velocities and line strengths are consistent with the observed H$_{2}$
emission lines in the $FUSE$ spectra.
We will focus on the lines in the $FUSE$ bandpass for the remainder of the paper, 
but we note that the
detection of additional H$_{2}$ fluorescent emission lines with an independent instrument
makes the conclusions presented in Sections 3 and 4 more robust.

\section{Analysis and Results} 

\subsection{Total Emitted H$_{2}$ Flux}

Using the measured H$_{2}$ emission line fluxes, we can calculate the total
fluorescent output from the \ion{O}{6} pumped cascade.  The total emitted flux
out of the electrovibrational state ($n',v',J'$),  $\sum\limits_{j} I_{ij}$, is given by
\begin{equation}
\sum\limits_{j} I_{ij}~=~I_{ik}~\left(~\frac{ A_{ik} }{ \sum\limits_{l} A_{il}}~\right)^{-1}~(1-\xi_{i})^{-1} 
\end{equation}
where $i$ refers to the upper state ($n'$,$v'$,$J'$).  The indices 
$j$, $k$, and $l$ refer to the lower states ($n''$,$v''$,$J''$).
$I_{ik}$ is the measured flux of the $FUSE$ band lines (in photons s$^{-1}$ cm$^{-2}$).
The ratios of individual to total Einstein $A$-values~\citep{abgrall93b} are the branching ratios, and $\xi_{i}$ is a correction
for the efficiency of predissociation in the excited electronic state~\citep{liu96}.  
In the case of AU Mic, we are concerned with Werner band emission 
($n'$~--~$n''$ = $C$~--~$X$~$\equiv$~$C^{1}\Pi_{u}$~--~$X^{1}\Sigma^{+}_{g}$), $v'$ = 1, and $v''$ = 3 and 4 for the 
1119.08 and 1163.81 \AA\ lines, 
respectively.  
The predissociation fraction for the Werner bands is zero 
($\xi_{C}$ = 0; Ajello et al., 1984).~\nocite{ajello84} 

Following this procedure, we arrived at the total emitted photon flux, 
$\sum\limits_{j} I_{ij}$,  
derived from the observed (1~--~3) 1119.08 \AA\ and (1~--~4) 1163.81 \AA\ emission lines
individually (4.3 $\pm$ 1.7 $\times$ 10$^{-4}$ and 3.1 $\pm$ 1.6 $\times$ 10$^{-4}$
photons s$^{-1}$ cm$^{-2}$ for the 
(1~--~3) and (1~--~4) lines, respectively.)  
For the determination of the total H$_{2}$ column density presented below, 
we take the average of these values,
$\sum\limits_{j} I_{ij}$ = 3.7 $\pm$ 2.3 $\times$ 10$^{-4}$ photons s$^{-1}$ cm$^{-2}$.  
The error on the total emitted flux was determined such that both the 
(1~--~3) 1119.08~\AA\ and (1~--~4) 1163.81~\AA\ values are consistent with the mean value; this approach ensures a conservative estimate of the measurement uncertainties.

\subsection{Inferred H$_{2}$ Column Density}

In order to determine the total column density that is associated with the observed level of
emission, we made three assumptions: 1) that fluorescence is the only source of the observed emission, 
2) that the stellar \ion{O}{6} $\lambda$1032 emission line is the only source of pumping photons, 
and 3) that the total number of photons is conserved.
The first assumption (1) seems warranted as H$_{2}$ cannot be electronically excited 
by shocks or collisional processes with other gas or dust particles. 
Additionally, excitation by electron collisions has a distinct far-UV emission signature~\citep{ajello82} 
which is not observed towards AU Mic.  
(2) AU Mic does not emit stellar continuum at wavelengths 
coincident with the absorbing transitions that produce the fluorescent 
emission lines, supporting the assumption that
the \ion{O}{6} emission line is responsible for the observed excitation.
(3) Detailed calculations of the formation and destruction of molecules in the 
AU Mic system are beyond the scope of this work, thus we assumed 
photon conservation for what follows.  

\begin{figure}
\begin{center}
\hspace{+0.0in}
\includegraphics[width=2.5in,angle=90]{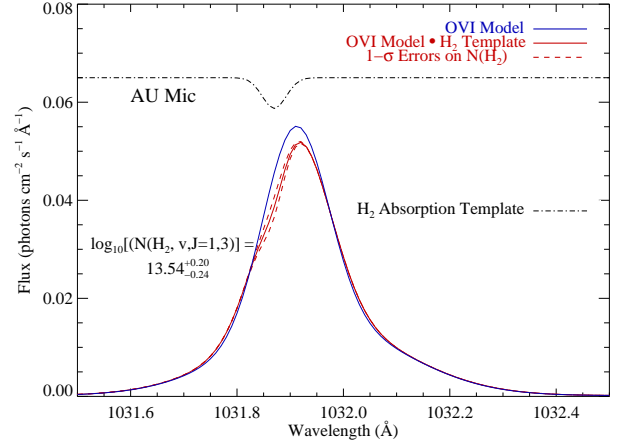}
\caption{\label{modabs} 
Excitation of fluorescent H$_{2}$ emission lines by stellar \ion{O}{6}  
emission.  The intrinsic stellar \ion{O}{6} $\lambda$1032 emission model (Roberge  
et al., 2005) is shown by the solid blue line.  The H$_{2}$ absorption  
model with a column density in the ($v$,$J$) = (1,3) level of 
$N_{X,1,3}$~=~3.48$^{+2.02}_{-1.48}$~$\times$~10$^{13}$~cm$^{-2}$  
is shown at the top with the black dot-dash line; this model has been  
convolved with the $FUSE$ LSF.  The solid red line shows the stellar  
emission model times the H$_{2}$ absorption model; the shape of the  
stellar emission line is modified by superimposed absorption from the H$_{2}$ 
$C$~--~$X$ (1~--~1) $Q(3)$ transition at 1031.87 \AA.  This absorption line is the  
excitation channel for the observed fluorescent H$_{2}$ emission lines  
shown in Figure 1.  The red dashed lines show total models with  
absorption columns corresponding to the quoted error ranges on the H$_{2}$  
column density $N_{X,1,3}$. 
}
\end{center}
\end{figure}

\citet{roberge05} modeled the \ion{O}{6} $\lambda\lambda$1032/1038 and 
\ion{C}{2} $\lambda\lambda$1036/1037 stellar emission lines in the $FUSE$ data by
fitting a combination of narrow and broad Gaussians to each line.  
The high-excitation ionic lines thought to originate in the chromosphere and 
transition regions of low mass dwarf
stars are known to be poorly fit by a single Gaussian component 
(Linsky \& Wood, 1994; Redfield et al., 2002). 
The narrow component of the profile has a FWHM$_{narrow}$ = 44 $\pm$ 6 km s$^{-1}$ with 
a velocity of v$_{narrow}$ = -4.6 $\pm$ 1.6 km s$^{-1}$.   This is consistent with the 
stellar velocity (v$_{*}$~=~-4.98 $\pm$ 0.02 km s$^{-1}$).  
The broad component has FWHM$_{broad}$ = 109 $\pm$ 25 km s$^{-1}$ with a somewhat 
redshifted velocity, v$_{broad}$ = +7.7 $\pm$ 9.3 km s$^{-1}$~\citep{roberge05}.  
This two-component model was used as the 
stellar emission profile to be absorbed by the $C$--$X$ (1~--~1) $Q(3)$ $\lambda$1031.87 \AA\ transition.
The absorption profile was created from an H$_{2}$ools optical depth template~\citep{h2ools} with a conservative 
Doppler parameter $b$ = 2 km s$^{-1}$~\citep{etangs01}.  
The H$_{2}$ools templates are optical depth arrays that can be used to fit arbitrary H$_{2}$ column
densities for many rovibrational states for $b$-values from 2~--~20 km s$^{-1}$~\citep{h2ools}.
The stellar emission model was binned to 0.01 \AA\ pixels to match the 
H$_{2}$ools wavelength grid.
A grid of column densities was searched to find the minimum difference in the absorbed 
[I(\ion{O}{6}$_{model}$)~--~I(\ion{O}{6}$_{model}$~$\times$~ H$_{2}$ absorption)] and emitted fluxes (the equilibrium condition).
This method found $N($H$_{2}$,$n,v,J)$ = $N_{X,1,3}$ = 3.48$^{+2.02}_{-1.48}$ $\times$ 10$^{13}$ cm$^{-2}$.
The \ion{O}{6} model and H$_{2}$ absorption are illustrated in Figure~\ref{modabs}.

The total $N($H$_{2}$$)$ can be determined from $N_{X,1,3}$ by assuming a
Boltzmann distribution and an estimate for the temperature.  ~\citet{liu04} assumed
that the CO in the AU Mic disk is in thermal equilibrium with the dust at 40 K and 
\citet{roberge05} consider absorption out of the $v$ = 0, $J$ = 0, 1 states of 
H$_{2}$ assuming that $T($H$_{2}$$)$~$\leq$~200K.  
To make our own determination of the H$_{2}$ excitation temperature, 
we used far-UV H$_{2}$ fluorescence models~\citep{france05a} 
to predict the temperature dependence of the H$_{2}$ emission spectrum.  
The relative fluxes of the emission lines are determined by the shape and  
strength of the exciting radiation field and the H$_{2}$ abundances in the 
rovibrational levels of the ground electronic state. 
The level populations are determined by the H$_{2}$ column  
density and excitation temperature.  
We computed fluorescence models for a range of
excitation temperatures (40 $\leq$ $T($H$_{2}$$)$ $\leq$ 2000 K) and 
column densities (10$^{16}$ $\leq$ $N($H$_{2}$$)$ $\leq$ 2 $\times$ 10$^{19}$).  
The fluorescence code used the 1030~--~1040~\AA\ \ion{O}{6}+\ion{C}{2} stellar 
emission model, described above, as the exciting radiation field.  

At excitation temperatures $\leq$ 700 K, fluorescent emission  
lines excited by absorption out of the $v$ = 0 and $J$ = 0, 1, and 2  
levels dominate the output in the 1100 - 1187~\AA\ wavelength range.   
These lines are not seen in the $FUSE$ spectra.		
The observed (1~--~3) and (1~--~4) emission lines become the strongest at temperatures  
above 800 K, providing a lower limit on the excitation temperature.   
This spectral variation with excitation temperature is due to the distribution
of higher rovibrational levels within the ground electronic state.  Once the
excitation temperature is high enough to significantly populate the ($v$,$J$) = (1,3) level,
this fluorescent route dominates due to the coincidence with \ion{O}{6} $\lambda$1032. 
An upper limit on $T($H$_{2}$$)$ can also be set from the observed spectral  
characteristics. The lack of Ly$\alpha$  pumped fluorescence indicates  
(\S 4.2) that the region producing H$_{2}$ line emission is cooler than  $\approx$ 2000 K.

\begin{deluxetable*}{ccccc}
\tabletypesize{\small}
\tablecaption{H$_{2}$-to-Dust Ratio Determinations for the AU Mic System. \label{gastodust}}
\tablewidth{0pt}
\tablehead{
\colhead{$M($H$_{2}$$)$} & \colhead{Technique} & \colhead{$M_{d}$\tablenotemark{a}} & 
\colhead{H$_{2}$-to-Dust} & \colhead{Reference} \\
($M_{\oplus}$) & & ($M_{\oplus}$) & &
}
\startdata
$\leq$~1.3	& 	CO (3~--~2) Emission	&	0.011	&	$\leq$~118:1	&	Liu et al., 2004	\\
$\leq$~0.07	& 	H$_{2}$ UV Absorption 	&	0.011	&	$\leq$~6:1	&	Roberge et al., 2005	\\
4.0~$\times$~10$^{-4}$	& 	H$_{2}$ Fluorescence & 0.011	&
0.036:1	&	This Work, T(H$_{2}$) = 800 K	\\
5.8~$\times$~10$^{-6}$ &	H$_{2}$ Fluorescence & 0.011	& 
5.2~$\times$~10$^{-4}$:1 & This Work, T(H$_{2}$) = 2000 K	\\
\enddata

\tablenotetext{a}{from 850 $\mu$m dust emission~\citep{liu04}}

\end{deluxetable*}

\begin{figure}[b]
\begin{center}
\hspace{+0.0in}
\includegraphics[width=2.6in,angle=90]{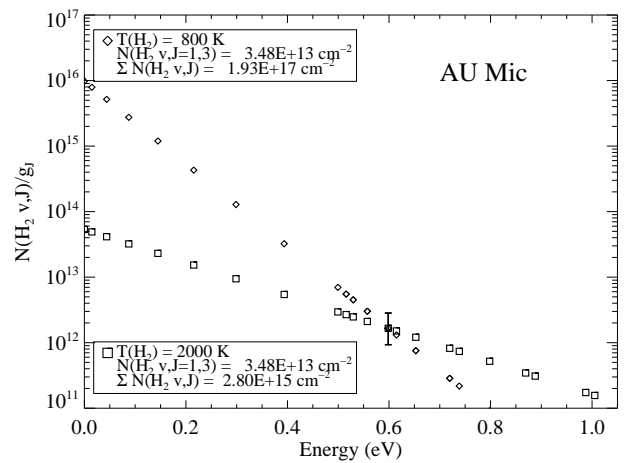}
\caption{\label{ncol} The H$_{2}$ column density distributions for
$T($H$_{2}$$)$~=~800 (diamonds) and 2000~K (squares).  
Excitation temperatures below 800 K are inconsistent with the observed
fluorescent spectrum.  Temperatures greater than 2000 K are ruled out by the 
absence of H$_{2}$ emission lines pumped by Ly$\alpha$.
Error bars on the column density are shown for the 
($v$,$J$) = (1,3) level. }
\end{center}
\end{figure}

In order to present a fiducial column density value that gives a sense of the measurement errors, 
we adopted $T($H$_{2}$$)$ = 1000 K as a characteristic temperature, 
which gives a total $N($H$_{2}$$)$ = 4.24$^{+2.46}_{-1.82}$ $\times$ 10$^{16}$ cm$^{-2}$.  
We emphasize however, [$T($H$_{2}$$)$,$N($H$_{2}$$)$] combinations ranging from 
[800 K, 1.93 $\times$ 10$^{17}$ cm$^{-2}$]
to [2000 K, 2.80 $\times$ 10$^{15}$ cm$^{-2}$] are consistent with the data
(see \S4.2 for a more detailed discussion of the molecular gas temperature).
Column density distributions are shown in Figure 3 for the 800 and 2000 K cases.



\section{Physical Properties of the Molecular Gas Component}

\subsection{Mass}
A dust mass ($M_{d}$) of 0.011 $M_{\oplus}$ in the AU Mic disk (at $d$~$\leq$ 70 AU) was measured 
by \citet{liu04} using 850 $\mu$m SCUBA observations.  
More recently, the dust mass in the AU Mic disk has been estimated from the visible and near-IR
scattered light profiles of the disk.  Calculations based on scattered light find masses smaller
($\sim$ a few~--~70 $\times$ 10$^{-4}$ $M_{\oplus}$ depending on the grain properties and size
distribution; Augereau \& Beust 2006) or equal ($\sim$ 0.01 $M_{\oplus}$; Strubbe \& Chiang 2006) to the 
sub-mm value.  In order to directly compare with H$_{2}$ absorption studies~\citep{roberge05}, we
will use 0.011 $M_{\oplus}$ as the CS dust mass here.  
\citet{liu04} also set an upper limit on 
the CO column density in the disk of $N($CO$)$~$\leq$~6.3~$\times$~10$^{13}$ cm$^{-2}$ from a CO (3~--~2)
346 GHz emission non-detection.  Assuming an $N($CO$)/N($H$_{2}$$)$ ratio of 10$^{-7}$, they 
place an upper limit on the H$_{2}$ column in the disk of $N($H$_{2}$$)$~$\leq$~6.3~$\times$~10$^{20}$ cm$^{-2}$.
Their assumption is supported by recent studies of the diffuse ISM at comparable 
values of $N($CO$)$ where the $N($CO$)/N($H$_{2}$$)$ ratio is observed to be in the 
range of a few $\times$~10$^{-7}$~\citep{burgh07}.  \citet{liu04} place an upper limit on the
mass of H$_{2}$ gas in the disk of $M_{H_{2}}$~$\leq$~1.3 $M_{\oplus}$ and the H$_{2}$-to-dust
ratio in the disk at $M_{H_{2}}/M_{d}$ $\leq$ 118:1.  
The limit on this ratio was 
further decreased by H$_{2}$ absorption line spectroscopy (Roberge et al. 2005; 
$N($H$_{2}$$)$~$\leq$~1.7~$\times$~10$^{19}$ cm$^{-2}$)
to $M_{H_{2}}/M_{d}$ $\leq$ 6:1.

Comparing the result derived in \S3.2 with that of~\citet{liu04}, we infer values for
$N($H$_{2}$$)$ that are $\approx$~3.3~$\times$~10$^{3}$~--~2.3~$\times$~10$^{5}$ below their upper limit.  The 
corresponding H$_{2}$-to-dust ratios are $M_{H_{2}}/M_{d}$ = (0.036~--~5.2~$\times$~10$^{-4}$):1, or 
$M_{H_{2}}/M_{d}$~$\lesssim$~$\frac{1}{30}$:1.  
This gives a total mass range of $M($H$_{2}$$)$ = 4.0~$\times$~10$^{-4}$~--~5.8~$\times$~10$^{-6}$~$M_{\oplus}$.
The value for $T($H$_{2}$$)$~=~1000 K is 
$M($H$_{2}$$)$ = 8.7$^{+5.2}_{-3.7}$~$\times$~10$^{-5}$~$M_{\oplus}$. 
These results are summarized in Table~\ref{gastodust}.

\subsection{Temperature}
Observations of H$_{2}$ emission excited by Ly$\alpha$ through the 
$B$~--~$X$ (1~--~2) $P(5)$ 1216.07 \AA\ and $B$~--~$X$ (1~--~2) $R(6)$ 1215.73 \AA\
coincidences are generally thought to indicate an H$_{2}$ ground state population 
characterized by temperatures $T($H$_{2}$$)$~$>$ 2000 K~\citep{black87,wood04,herczeg04,lupu06}.
In the case where \ion{O}{6} and Ly$\alpha$ excitation are both observed, 
Ly$\alpha$ excitation usually dominates (e.g.- T Tauri stars; 
Wilkinson et al. 2002; Herczeg et al. 2005).~\nocite{wilkinson02,herczeg05}
The observation of \ion{O}{6} pumped fluorescence in conjunction with the 
absence of Ly$\alpha$ fluorescence allows us to constrain
the molecular gas temperature 
in the AU Mic disk.  As discussed in \S3.2, the observed fluorescence spectrum
sets the lower limit on $T($H$_{2}$$)$~$\geq$ 800 K, and we suggest that the lack
of Ly$\alpha$ pumped lines in the $FUSE$ 
data
imply an upper limit of $T($H$_{2}$$)$~$<$~2000 K.

A quantitative calculation of the expected flux from Ly$\alpha$ induced fluorescence
is complicated by strong interstellar \ion{H}{1} absorption of the
line profile~\citep{pagano00,redfield02}. It seems clear that the 
local AU Mic Ly$\alpha$ radiation field is at least an order of magnitude more intense
than the local \ion{O}{6} radiation field (Figure 5 in both Pagano et al. 2000 and 
Redfield et al. 2002).  The $A$-values for the strongest
Ly$\alpha$ pumped lines are similar (to within 30 \%) to those 
pumped by \ion{O}{6}~\citep{abgrall93a,abgrall93b}. 
Ignoring optical depth and extinction effects, the column densities in the 
absorbing transitions control the resultant emission spectrum. 
We can define the ratio of column densities in the 
relevant states $R^{OVI}_{Ly\alpha}$~$\equiv$~$N(1,3) / (N(2,5) + N(2,6))$.
When $R^{OVI}_{Ly\alpha}$~$\lesssim$~10 (this value is set by the rough estimate
of stellar Ly$\alpha$-to-\ion{O}{6}), we would expect a detectable contribution
from Ly$\alpha$ pumped fluorescence.  For gas temperatures of 
$T($H$_{2}$$)$ = [1000, 2000, 3000, 4000 K], the corresponding ratios are
$R^{OVI}_{Ly\alpha}$ = [545.4, 16.0, 4.9, 2.7].  Thus, only when
$T($H$_{2}$$)$~$<$~2000 K do we expect to detect H$_{2}$ emission from 
\ion{O}{6} excitation in the absence of lines excited by Ly$\alpha$.

\subsection{Spatial Origin and Heating}
The H$_{2}$ required to produce the observed fluorescence need not be
coincident with the stellar line of sight.  The emitting gas could
reside in a cloud that extends beyond the disk.  If the necessary 
column density suggests absorption that is not observed along the line of sight, 
this could be evidence for an extra-planar gas component. In order to explore this possibility, 
we compared a model of the \ion{O}{6} line profile modified by the required H$_{2}$ 
absorption to the \ion{O}{6}  
profile observed in the FUSE spectra (shown in Figure 4).  This  
assumes that the total required absorbing column density lies along  
the line of sight, as it would if the H$_{2}$ gas is entirely in the  
disk.  The \ion{O}{6} model with the required H$_{2}$ absorption superimposed is  
consistent with the data, within the measurement uncertainties.  This  
leaves open the possibility that all of the emitting H$_{2}$ gas lies in  
the disk, and only the relatively small absorbing column density and  
low signal-to-noise of the FUSE spectra prevent it from being clearly  
detected in absorption against the \ion{O}{6} 1032 emission line, as has been 
observed in other CS disks~\citep{roberge01}.
It is interesting to note that the column density range we derive here
($N($H$_{2}$$)$~=~1.9 $\times$ 10$^{17}$~--~2.8 $\times$ 10$^{15}$ cm$^{-2}$) 
is consistent with the possible line of sight absorption suggested
in Section 4.2 of~\citet{roberge05}.

\begin{figure}
\begin{center}
\hspace{+0.0in}
\includegraphics[width=2.5in,angle=90]{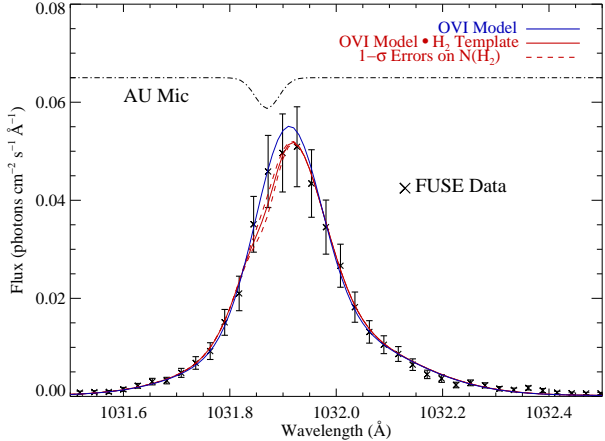}
\caption{\label{o6check} 
A comparison between the predicted and observed \ion{O}{6} line profiles.   
The blue and red lines are the same as in Figure 2.  The FUSE data  
are overplotted with x's.  The required H$_{2}$ absorption is consistent  
with the observed line profile, within the measurement errors.  This  
indicates that the  H$_{2}$ absorption could arise completely within the AU  
Mic CS disk.
}
\end{center}
\end{figure}

We can also test if our $T($H$_{2}$$)$ estimate is consistent
with a disk origin for the emitting gas.  
$T($H$_{2}$$)$ $\sim$ 1000 K is warmer than 
typical debris disk gas temperatures considered in previous theoretical
($T$ $\leq$ 300K, Kamp \& van Zadelhoff 2001) and observational
($T$~$\sim$~40 K, Liu et al. 2004; $\leq$ 200 K, Roberge et al. 2005)  
studies.~\nocite{kamp01}  The first possibility is that 
the gas is in close proximity to the star.  Calculations
have been presented for the detectability of Ly$\alpha$ 
pumped H$_{2}$ in the spectra of late-type stars by~\citet{jordan78}, 
but as we have shown, the temperature of the molecular gas associated
with AU Mic is below the level required for that process.  Starspots have
lower effective temperatures than the conventional photosphere, and the 
AU Mic spot temperature has been measured in variability studies~\citep{rodono86}.
However, the observed spot temperature, $T_{spot}$~=~2650 K, is still substantially
above our 2000 K limit.  The ratio of spot-to-unspotted temperatures in AU Mic is $\sim$~0.76. 
This value agrees well with the spot-to-unspotted temperature ratios found for more massive active 
stars ($\sim$~0.66~--~0.86;
Neff et al. 1995; O'Neal et al. 1996) from an analysis of TiO absorption bands.~\nocite{neff95,oneal96}
We find that even in the coolest regions of the AU Mic surface, 
a photospheric origin for the observed H$_{2}$ emission can be 
most likely ruled out. 

The AU Mic corona is characterized by electron temperatures
in the range of 10$^{4}$~$\leq$~$T_{e}$~$\leq$~10$^{7}$ K~\citep{maran94}.
Electron densities in the transition region and coronal regime are
of order $n_{e}$ = 5~$\times$~10$^{10}$ cm$^{-3}$~\citep{maran94}, 
with suggestions of values several orders of magnitude greater~\citep{redfield02}.
We can estimate the characteristic survival time for molecules near the 
AU Mic transition and coronal regions 
by assuming that the H$_{2}$ is at low densities (i.e.- not contained in 
dense clumps), using $e^{-}$ + H$_{2}$ cross sections to determine the 
collisional dissociation rate~\citep{martin98}.  Taking a conservative value for the 
electron temperature (10$^{4}$ K), we find the e-folding time for H$_{2}$ 
dissociation via electron impact, $\tau_{diss}$ = ($\gamma_{e} n_{e}$)$^{-1}$, 
where $\gamma_{e}$ is the electron impact dissociation rate coefficient, 
to be $\tau_{diss}$~$\leq$ 1 minute.  
It is possible that the interstellar conditions assumed to estimate this 
time scale do not apply directly to the AU Mic environment, 
however the basic picture that 
molecules cannot survive in the immediate environment of AU Mic seems
robust even if the parameters vary by several orders of magnitude.
As an additional constraint, we consider the possible observational
consequences assuming that molecules could survive in regions as close as a few
stellar radii. H$_{2}$ heated by $e^{-}$ collisions in the stellar atmosphere 
might be expected to show the well characterized signature of electron-impact induced
excitation~\citep{ajello82,ajello84}.  The electron-impact
excitation spectrum of H$_{2}$ peaks in the $FUSE$ and STIS bandpasses~\citep{gustin04,gustin06}, and no
emission from these features is observed.

Some grain species may be resistant to sublimation at $r$~$<$ 6 $R_{*}$~\citep{mann06},
and it may be possible that the H$_{2}$ is heated by collisions with 
grains in thermal equilibrium with the stellar radiation field 
($r$~$\approx$ 6 $R_{*}$ for $T$ = 1000K). 
However, previous studies have found no dust emission or scattered light in the 
inner disk near AU Mic.
In the optical, the disk is cleared inside $r$ $\leq$~7.5 AU~\citep{krist05}. 
Probing the 850 $\mu$m dust emission, the inferred inner radius is $r$~$\leq$~17 AU~\citep{liu04}.
~\nocite{krist05,liu04}

The inferred range of $T($H$_{2}$$)$ could also be produced by heating
processes operating in the disk.  There have been considerable efforts towards 
modeling the gas and dust components of CS disks
(Kamp \& van Zadelhoff 2001; Jonkheid et al. 2004; Besla \& Wu 2007; and
references therein),~\nocite{kamp01,jonkheid04,besla07} although most
of these efforts have focused on higher mass stars (Herbig Ae and T Tauri stars).
Nevertheless, we will use these models to understand the important processes
heating the molecular gas in AU Mic, noting where certain assumptions are invalid for 
an M star disk.  Heating processes include photoelectric heating by grains, collisional
de-excitation of H$_{2}$, photodissociation of H$_{2}$, H$_{2}$ formation, gas-grain collisions, 
carbon ionization, gas-grain drift, and cosmic ray heating~\citep{kamp01}.
Relevant cooling processes that regulate the gas temperature include
[\ion{O}{1}] cooling (from $\lambda$6300 \AA\ and $\lambda$63.2 $\mu$m), [\ion{C}{1}] and [\ion{C}{2}] cooling, H$_{2}$ rovibrational line cooling, 
Ly$\alpha$ cooling, and CO cooling~\citep{kamp01}. 

Gas-grain collisions only heat the gas when $T_{gas}$ $<$ $T_{dust}$, 
and there is uncertainty whether gas-grain drift is an efficient heating mechanism~\citep{besla07}.
The gas temperature would have to be much higher than we observe for 
Ly$\alpha$ or [\ion{O}{1}] $\lambda$6300 \AA\ cooling to contribute significantly, and the 
lifetimes of the rovibrational states of H$_{2}$ are very long, making 
these IR transitions inefficient coolants.  
CO emission is not detected~\citep{liu04}, so it is hard to assess this
contribution to the cooling.  Finally, AU Mic lacks the far-UV
(912~--~1110 \AA) stellar continuum that drives the photodissociation of H$_{2}$ and 
CO, and produces \ion{C}{2} through the photoionization of carbon. 
Processes that depend on this flux ($heating:$ photodissociation of H$_{2}$ and 
carbon ionization, $cooling:$ [\ion{C}{2}] emission)
must be driven solely by the interstellar radiation field~\citep{draine78}. 
Relative to models of A-star disks, we presume these processes are 
of diminished importance in AU Mic.     

This leaves photoelectric and H$_{2}$ formation heating, and far-IR fine structure
line cooling as the dominant processes that determine the gas temperature in the AU Mic 
disk.  This conclusion generally agrees with the scenario put forth by~\citet{besla07}, who
calculate gas temperatures in the disks of more massive stars (K2 and earlier) of 
T$_{gas}$ $>$ 400 K, depending on the model parameters.  
\citet{jonkheid04} find 
that strong photoelectric heating leads to gas temperatures as high as 1000 K in the 
surface regions of disks, although they consider more strongly flared disks than AU Mic.  
We note that 1000 K is similar to the temperature found for interstellar
H$_{2}$ where grain-formation pumping is thought to be a dominant excitation source~\citep{spitzer73}.
Suffice to say, photoelectric and H$_{2}$ formation heating seem to be capable of 
elevating the gas temperature to roughly the observed level, 
but more modeling work is needed for disks around low mass stars.
Combining this with the arguments against a stellar origin for the observed H$_{2}$ emission
given above, we favor the hypothesis that the fluorescent emission originates in the 
CS disk.
~\nocite{jordan78,kamp01,besla07}

\section{Comparison with $\beta$ Pic}

AU Mic is a member of the $\beta$ Pictoris moving group~\citep{zuckerman01}, meaning the AU Mic
disk is roughly the same age as the well-studied $\beta$ Pic debris disk.  $\beta$ Pic (A5V)
is roughly 3.6 times more massive than AU Mic~\citep{kalas04}.  In this section, we briefly compare the 
molecular gas properties derived for AU Mic with previous observations of gas in the $\beta$ Pic CS disk.

\begin{deluxetable*}{lccc}
\tabletypesize{\small}
\tablecaption{Predicted Near and Mid-IR H$_{2}$ AU Mic Emission Line Brightnesses. \label{irs_pred}}
\tablewidth{0pt}
\tablehead{
\colhead{Line} & \colhead{Wavelength} & \colhead{Instrument}   
& \colhead{Brightness\tablenotemark{a}} \\ 
($v'$~--~$v''$) $S(J)$ & ($\mu$m) &  & (ergs cm$^{-2}$ s$^{-1}$ sr$^{-1}$)
}
\startdata
(1~--~0) $S(1)$	& 2.12 & $HST$-NICMOS/Ground 	& 9.01 $\times$ 10$^{-7}$ \\
(2~--~1) $S(1)$	& 2.25 & Ground 		& 1.13 $\times$ 10$^{-9}$~--~7.47 $\times$ 10$^{-8}$ \\
(0~--~0) $S(7)$	& 5.51 & $Spitzer$-IRS 		& 5.74 $\times$ 10$^{-7}$~--~4.93 $\times$ 10$^{-7}$ \\
(0~--~0) $S(6)$	& 6.11 & $Spitzer$-IRS 		& 2.95 $\times$ 10$^{-7}$~--~1.48 $\times$ 10$^{-7}$ \\
(0~--~0) $S(5)$	& 6.91 & $Spitzer$-IRS 		& 2.02 $\times$ 10$^{-6}$~--~3.30 $\times$ 10$^{-7}$ \\
(0~--~0) $S(4)$	& 8.03 & $Spitzer$-IRS/Ground 	& 8.84 $\times$ 10$^{-7}$~--~6.40 $\times$ 10$^{-8}$ \\
(0~--~0) $S(3)$	& 9.66 & $Spitzer$-IRS 		& 2.29 $\times$ 10$^{-6}$~--~8.13 $\times$ 10$^{-8}$ \\
(0~--~0) $S(2)$	& 12.28 & $Spitzer$-IRS/Ground 	& 3.84 $\times$ 10$^{-7}$~--~7.39 $\times$ 10$^{-9}$ \\
(0~--~0) $S(1)$	& 17.03 & $Spitzer$-IRS/Ground 	& 2.55 $\times$ 10$^{-7}$~--~2.99 $\times$ 10$^{-9}$ \\
(0~--~0) $S(0)$ & 28.22 & $Spitzer$-IRS 	& 4.25 $\times$ 10$^{-9}$~--~3.41 $\times$ 10$^{-11}$ \\
 \enddata


\tablenotetext{a}{Predicted brightness ranges reflect the adopted temperature ranges of
$T($H$_{2}$$)$ = 800~--~2000 K, corresponding to $N($H$_{2}$$)$ = 
1.9 $\times$ 10$^{17}$ ~--~ 2.8 $\times$ 10$^{15}$ cm$^{-2}$, 
described in \S3.2.}
\end{deluxetable*}

Multi-wavelength model fits to the $\beta$ Pic CS disk SED predict a dust mass of 
0.037 $M_{\oplus}$~\citep{dent00}.  The molecular hydrogen mass in $\beta$ Pic is 
less clearly defined.  $ISO$ observations of the mid-IR emission lines of H$_{2}$ find
a large molecular gas reservoir ($M($H$_{2}$$)$ $\approx$ 57 $M_{\oplus}$) 
associated with $\beta$ Pic~\citep{thi01b}.  
It seems unlikely that this emission is distributed uniformly throughout the disk.
Using UV absorption techniques analogous
to those presented for AU Mic~\citep{roberge05},~\citet{etangs01} report a non-detection
of H$_{2}$ absorption in the edge-on disk.  They set an upper limit on the H$_{2}$
column density of $N($H$_{2}$$)$~$\leq$~10$^{18}$ cm$^{-2}$, corresponding to a molecular 
gas mass of $\leq$~0.095 $M_{\oplus}$.

The ratio of the $\beta$ Pic and AU Mic dust masses ($\sim$ 3.4) is approximately equal to the 
ratio of their stellar masses ($\sim$ 3.6).    The observed dust in both disks is thought
to be continually replenished.  Collisions of larger parent bodies in the disk
can repopulate the small grain population that is detected as scattered light in the 
optical/near-IR and as thermal emission at longer wavelengths.  Models have shown that
this scenario can reproduce the observed dust properties of $\beta$ Pic~\citep{thebault03}
and AU Mic~\citep{augereau06,strubbe06}.  Grain collisions may also be responsible for  
replenishing the metallic gas observed in the $\beta$ Pic disk~\citep{fernandez06},
but it is unclear if this process contributes to the gas phase H$_{2}$ abundance. 

The H$_{2}$-to-dust ratio in the $\beta$ Pic disk 
(from the UV absorption line upper limit) is $<$~3:1.  This is consistent
with our range of H$_{2}$-to-dust ratios for AU Mic (0.036~--~5.2~$\times$~10$^{-4}$:1).  
If H$_{2}$ gas is present in the $\beta$ Pic disk at a similar gas-to-dust ratio as AU Mic, 
a natural question arises: Why were the fluorescent emission lines not detected in 
$FUSE$ observations of $\beta$ Pic?  The answer is most likely related to the earlier 
spectral type of $\beta$ Pic.  The stellar \ion{O}{6} emission line from
an active M-star such as AU Mic is considerably stronger than in $\beta$ Pic.  The 
peak flux at the \ion{O}{6} $\lambda$1032~\AA\ line center (coincident
with the absorbing H$_{2}$ transition studied here) is over an order of magnitude higher
in AU Mic.  Additionally, the stellar continuum of $\beta$ Pic extends down to 
1100 \AA~\citep{etangs01}, lowering the 
line-to-continuum ratio
at the strongest emission line wavelengths.

\section{IR Brightness Predictions}

H$_{2}$ does not have an intrinsic dipole moment, hence the rovibrational transitions
of the molecule proceed by the slower quadrupole channel, making them
optically thin in most astrophysical environments~\citep{black87}.  Assuming the optically thin case, we used the
derived column density distributions 
([$T($H$_{2}$$)$,$N($H$_{2}$$)$] = [800 K, 1.93 $\times$ 10$^{17}$ cm$^{-2}$]~--
[2000 K, 2.80 $\times$ 10$^{15}$ cm$^{-2}$])
to predict the near and mid-IR H$_{2}$ emission line strengths 
for transitions that are observable from ground-based 
facilities~\citep{speck03,allers05}, $HST$-NICMOS~\citep{meixner05}, or $Spitzer$-IRS~\citep{irs04,hora06}.
The most readily observable H$_{2}$ lines are the 
rovibrational lines (1~--~0) $S(1)$ $\lambda$2.12 $\mu$m and (2~--~1) $S(1)$ $\lambda$2.25 $\mu$m, 
and the pure rotational lines (0~--~0) $S(7)$~--~$S(0)$ $\lambda$5~--~29 $\mu$m.
The predicted 
line intensity can be calculated from 
\begin{equation}
I_{mS(J'')} = \left(\frac{N_{p(J''+2)}~A_{q}}{\lambda_{q}}\right)
\frac{hc}{4\pi}
\end{equation}
where $I_{mS(J")}$ is in units of ergs s$^{-1}$ cm$^{-2}$ sr$^{-1}$~\citep{black87,rosenthal00}. $m$ is the ($v'$~--~$v''$) transition, $p$ refers to the upper vibrational level, 
and $q$ labels the $A$-value and wavelength for the relevant transition.
The $J''+2$ notation is a consequence of the $S$ branch ($\Delta J$ = +2) transition. 
The $A$-values are from \citet{wolniewicz98}.
The predicted line strengths are given in Table 3.  

The exact detection limits of these lines will be determined by the angular size of the AU Mic 
emission, a larger filling fraction will increase the observed signal at a given surface 
brightness.  The $Spitzer$-IRS has access to the largest number of these lines. 
Even assuming optimistic aperture filling fractions of unity, the brightest of these lines are predicted
to have F$_{\nu}$~$<$ 0.05 mJy, which is at or below the IRS noise limit.  
This result is consistent with the non-detection of H$_{2}$
emission from the 15 CS disks around young Sun-like stars in the 
Formation and Evolution of Planetary Systems $Spitzer$ Legacy Program~\citep{pascucci06}.
$HST$-NICMOS has the capability for narrow band imaging in the (1~--~0) 
$S(1)$ $\lambda$2.12 $\mu$m line, however we predict that achieving
contrast with the stellar emission in this bright star ($m_{V}$~=~8.8; Kalas et al. 2004)
will be difficult. At a temperature of $T_{eff}$ = 3500 K, the photosphere will emit
strongly in the near-IR ($\lambda_{max}$~=~0.83 $\mu$m). 
Near-IR imaging will require high dynamic range in flux to 
achieve contrast between the stellar emission and the faint molecular gas.
Calculations for other instruments can be performed using the values in Table 3.
We suggest that high resolution near-IR spectroscopy could be the most promising 
technique for future observations of H$_{2}$ in the disk.  


\section{Summary}

We have presented far-UV observations of H$_{2}$ emission in the AU Microscopii 
CS disk.  The spectra displayed fluorescent emission lines
excited by stellar \ion{O}{6} $\lambda$1032 photons coincident with the 
$C$--$X$ (1~--~1) $Q(3)$ $\lambda$1031.87 \AA\ transition.  These lines
imply a total column density in the molecular gas of 
$N($H$_{2}$$)$~=~1.9 $\times$ 10$^{17}$~--~2.8 $\times$ 10$^{15}$ cm$^{-2}$.
This detection is roughly two orders of magnitude smaller than published upper limits
on $N($H$_{2}$$)$.  Comparing this value with previous limits on the 
gas mass in the system, we find 
$M($H$_{2}$$)$ = 4.0~$\times$~10$^{-4}$~--~5.8~$\times$~10$^{-6}$ $M_{\oplus}$.
Using the molecular mass and the 850 $\mu$m dust emission, we found
a gas-to-dust ratio of $\lesssim$~$\frac{1}{30}$:1.  The derived column densities
and gas masses depend upon the assumed excitation temperature, which we estimate
to be 800~--~2000 K.  We presented the basis for this temperature distribution, and 
discussed the value in the context of the AU Mic system.  We conclude that the
warm H$_{2}$ is most likely associated with the disk, with photoelectric heating
and formation pumping as the dominant heating mechanisms.  The 
intensity was predicted for several near and mid-IR
lines of H$_{2}$.  These lines are accessible to current ground and space-based 
observatories, although due to the low column density and weak 
intrinsic nature of the lines, they will be challenging to detect.

\acknowledgments
We thank Alexis Brandeker, Yanqin Wu, Ray Jayawardhana, and Joerg Fischera for 
enjoyable discussions about circumstellar disk structure and Leslie Hebb for a discussion
of M-star surface activity.  
We also appreciate insightful input from Stephan McCandliss 
and Peter Martin on photon and electron processes involving H$_{2}$.  
S. R. acknowledges support provided by NASA through Hubble Fellowship grant 
HST-HF-01190.01 awarded by the Space Telescope Science Institute, which is 
operated by the Association of Universities for Research in Astronomy, Inc., 
for NASA, under contract NAS 5-26555.
The $FUSE$ data were obtained as part of the 
NASA-CNES-CSA $FUSE$ mission, operated by The Johns Hopkins University.



\bibliography{ms_em}




\clearpage

\end{document}